\documentclass[aps,prd,preprintnumbers,nofootinbib,superscriptaddress,preprint]{revtex4-1}
\usepackage{graphicx}
\usepackage{amssymb}
\usepackage{amsmath}
\usepackage{color}
\usepackage{booktabs}
\usepackage{dcolumn}
\usepackage{ulem}

\graphicspath{{fig/}}

\newcommand{\beq}{\begin{eqnarray}}
\newcommand{\eeq}{\end{eqnarray}}

\def\fsl#1{\setbox0=\hbox{$#1$}
   \dimen0=\wd0
   \setbox1=\hbox{/} \dimen1=\wd1
   \ifdim\dimen0>\dimen1
      \rlap{\hbox to \dimen0{\hfil/\hfil}}
      #1
   \else
      \rlap{\hbox to \dimen1{\hfil$#1$\hfil}}
      /
   \fi}

\begin{document}

\preprint{UTHEP-720, UTCCS-P-112}

\title{Reply to ``Comment on `Relation between scattering amplitude and 
Bethe-Salpeter wave function in quantum field theory'\!\! ''}

\date{\today
}

\author{Takeshi~Yamazaki}
\affiliation{Faculty of Pure and Applied Sciences, University of Tsukuba, \\ Tsukuba, Ibaraki 305-8571, Japan}
\affiliation{Center for Computational Sciences, University of Tsukuba, \\ Tsukuba, Ibaraki 305-8577, Japan}
\affiliation{RIKEN Advanced Institute for Computational Science,
Kobe, Hyogo 650-0047, Japan}
\author{Yoshinobu Kuramashi}
\affiliation{Center for Computational Sciences, University of Tsukuba, \\ Tsukuba, Ibaraki 305-8577, Japan}
\affiliation{RIKEN Advanced Institute for Computational Science,
Kobe, Hyogo 650-0047, Japan}

\begin{abstract}

We reemphasize the momentum dependence of the coefficients
of the derivative expansion as already explained in our paper~\cite{Yamazaki:2017gjl}.
We also discuss how the momentum dependence plagues the
time-dependent HALQCD method and what is 
a necessary condition for the method
to yield valid results being independent of the choice of 
the interpolating operators.

\end{abstract}

\maketitle

Once again we stress the momentum dependence of the
coefficients in the derivative expansion as already explained in Ref.~\cite{Yamazaki:2017gjl}
and how the momentum dependence affects the time-dependent 
HALQCD method~\cite{HALQCD:2012aa}\footnote{The time-dependent HALQCD method is a procedure in the HALQCD method to determine the coefficients in the derivative expansion without specifying the input momenta as explained in Appendix.}.
We also show a necessary condition to
obtain valid results from the method, which should be independent of the choice of the interpolating operators in the source time slice.
Finally we add some remarks on the various methods which have been proposed so far to calculate the scattering amplitude.

First of all, $k$ dependence of the expansion coefficient $V_i(x)$
of the reduced BS wave function $h(x;k)$ is
inevitable in a practical determination by lattice calculations, as far as 
the expansion is truncated at some finite order~\cite{Yamazaki:2017gjl}.
This is true even if $V_i(x)$ is defined in a $k$ independent way
in the derivative expansion with the infinite terms 
as shown in Ref.~\cite{Aoki:2017yru}.
There is no contradiction between the above two points.
Actually, the authors in Ref.~\cite{Aoki:2017yru} admit that
``{\it In more realistic cases with higher order derivative terms in Eq.(2), the phase shift $\delta(k)$ calculated from $V_{0,2}$ as constructed in Eqs.(6) is exact at ${\bf q} = {\bf k}$ and ${\bf k}^\prime$, and is only approximate at other {\bf q}.}''
In other words, the coefficients are varied when the input momenta\footnote{The input momentum is defined by the energy of interacting two particles in the finite box so that it is not known a priori and should be measured in lattice calculations.},
${\bf k}$ and ${\bf k}^\prime$ in this case, are changed.
This is exactly a statement that the coefficients depend on the input momenta used in the practical determination.

Related to the above issue, we comment on another statement in Ref.~\cite{Aoki:2017yru}:
``{\it The primary confusion of Ref.~[1] originates from a claim that $V ({\bf r}; {\bf k})$ in Eq.(1) is replaced by $V ({\bf r}; {\bf q})$ even for ${\bf q} \ne {\bf k}$ in the HAL QCD method. Such a replacement however has never been introduced in the HAL QCD method.}''
In our paper~\cite{Yamazaki:2017gjl} we cannot find such a claim that $V ({\bf r}; {\bf k})$ is replaced by $V ({\bf r}; {\bf q})$ even for ${\bf q} \ne {\bf k}$ in the HALQCD method.
What we pointed out in Ref.~\cite{Yamazaki:2017gjl} is abuse of the potential determined by the HALQCD method. In the leading order HALQCD method, for example, 
$h(x;k)$ is approximated by only one term $V_0(x)$ and the scattering phase shifts for the wide range of the momentum region are 
presented by solving the Schr\"odinger equation with $V_0(x)$.
In principle, however, this procedure can afford to give a correct result only at the input momentum used in the determination of the potential, while
it is not so in other momenta.
To make matters worse, the input momentum, where the correct results should be obtained, cannot be determined from the time-dependent HALQCD method
as discussed below. So we never know at which momentum the scattering phase shift given by the HALQCD method is correct.

In Ref.~\cite{Aoki:2017yru} there is a statement on
the time-dependent HALQCD method~\cite{HALQCD:2012aa}:
``{\it In practice, the time-dependent HAL QCD method based on the Euclidean-time ($t$) dependence of the hadronic correlation function is a useful equivalent method to treat those states with different
momentums simultaneously, as demonstrated in [7].
}''
We also need to comment on the method,
because it is based on the assumption that $V_i(x)$ is independent of $k$.
In Appendix~\ref{app:thal},
we explain how the assumption is used in the method
and what is a necessary condition to obtain a valid result of $V_i(x)$
in a practical lattice calculation, where the expansion of $h(x;k)$ is 
truncated at some finite terms, {\it i.e.}, $V_i(x)$ depends on $k$.
The valid result means that it should be independent of the choice of the interpolating source operators 
in the correlation functions.
This condition has never been discussed in 
all the calculations using the method, 
see Ref.~\cite{Iritani:2018zbt} for example.

In order to obtain valid $V_i(x)$ with the truncated expansion, 
the effective number of elastic scattering states contributing to 
the correlation functions must be the same as the number of the operators,
{\it i.e.}, the number of $V_i(x)$ in the expansion.
It is exactly the same condition to obtain the energy from a
correlation function matrix using
the generalized eigenvalue problem~\cite{Luscher:1990ck}.
Contrary to the claim of Ref.~\cite{Iritani:2018zbt}, this condition is generally not satisfied in a region where the
inelastic scattering state contributions start to become negligible
in the correlation functions.
If the condition is satisfied, 
the time-dependent HALQCD method is allowed to give the valid $V_i(x)$ only
at the momenta of the states in the correlation functions, whose number should be the same as the number of $V_i(x)$.
In this case, however, it is a critical defect that the method cannot specify the momenta where $V_i(x)$ gives correct scattering amplitudes, because
the values of the momenta cannot be determined in
the method.

We also comment on
a statement in Ref.~\cite{Aoki:2017yru} that
``{\it 
In Ref. [1], there is also a statement that ``Therefore, a smearing of the interpolating operator in the BS wave function gives a different scattering amplitude from the one obtained from the fundamental relation, which depends on the smearing function s(x).''.
As already shown explicitly in Sec.II.D of [8], this statement is mathematically incorrect.}''.
What we have shown in Ref.~\cite{Yamazaki:2017gjl} is that the scattering amplitude with the smearing interpolating operator\footnote{This corresponds to a smearing of the sink operator in the correlation functions in the above discussion.} $\tilde{H}(k;k)$ depends on the smearing function, and
differs from the one with the local interpolating operator $H(k;k)$.
The statement in Ref.~\cite{Yamazaki:2017gjl} is absolutely correct in the mathematical sense, because
the difference between $\tilde{H}(k;k)$ and $H(k;k)$ is explained by 
the overall factor depending on the
smearing function and the momentum $k$.

Finally, we comment on the various methods to calculate the scattering amplitude, which have been proposed so far.

The finite volume method using the formula connecting the scattering amplitude and the momentum~\cite{Luscher:1986pf,Luscher:1990ux} 
is the best approach to calculate the amplitude in the sense that
it is obtained from only the input momentum 
determined from the two-particle energy in a finite box.
The formula is derived from the BS wave function
outside the interaction range~\cite{Lin:2001ek,Aoki:2005uf}.
CP-PACS Collaboration has shown that its $x$ dependence gives
the consistent momentum with
the one determined from the two-particle energy~\cite{Aoki:2005uf}.
This is a confirmation that the BS wave function
outside the interaction range can be used to obtain the scattering amplitude.
In another approach with the use of information of
the BS wave function inside the interaction range,
{\it i.e.}, the reduce BS wave function $h(x;k)$~\cite{Yamazaki:2017gjl},
the scattering amplitude can be directly obtained through
a simple formula called the fundamental relation~\cite{Yamazaki:2017gjl},
\begin{equation}
\frac{4\pi}{k}e^{i\delta(k)}\sin\delta(k) =
-\int d^3x h(x;k) e^{-i\vec{k}\cdot\vec{x}} ,
\label{eq:frelation}
\end{equation}
without any assumptions.
The first lattice calculation using this relation~\cite{Namekawa:2017sxs},
has shown that the scattering length calculated from the above relation
agrees with the one from the finite volume method.
A drawback in this approach is that we need the BS wave function in addition to the input momentum.
On the other hand, it may be advantageous that this approach is not based on the formula connecting the scattering amplitude and the momentum: 
Calculation of the scattering amplitude
of more than two particles might be easier if we can find a similar relation
corresponding to the fundamental relation of Eq.~(\ref{eq:frelation}).
Since it was derived through
the LSZ reduction formula, its extension to more than two particles could be
straightforward.

The HALQCD method is also classified into the second approach.
However, the procedure to obtain the scattering amplitude is redundantly complicated than the direct method proposed in Ref.~\cite{Yamazaki:2017gjl}.
The HALQCD method first determines the potential from the BS wave function inside the interaction range by fitting the data with some assumption of the potential form and solves the Schr\"odinger equation with the potential as input. Then, the scattering amplitude is determined from the wave function obtained from the Schr\"odinger equation.
The complexity of the method introduces additional serious systematic uncertainties such as the convergence of the derivative expansion and inability of specifying momenta where the correct scattering amplitudes should be obtained.
It is not clear whether the expansion converges or not
by investigating the convergence properties with 
only a few terms as in Ref.~\cite{Iritani:2018zbt},
because it is not a systematic expansion.
Furthermore, it could be possible that
the necessary condition for the time-dependent HALQCD method is not 
satisfied as discussed above.
The magnitude of these systematic errors are hardly estimated by 
the HALQCD method itself, which is a typical feature of model calculation, so that the results should be always checked by other methods.

\appendix

\section{time-dependent HALQCD method}
\label{app:thal}

The time-dependent HALQCD method~\cite{HALQCD:2012aa} is claimed to
obtain $k$ independent $V_i(x)$ by solving simultaneous equations
of two-particle correlation functions on the lattice.
In this appendix we discuss that such $V_i(x)$ cannot be obtained
in a practical calculation,
and a condition to obtain valid results, which do not depend on
the choice of the interpolating operators in the source time slice.

The correlation function of the two pions on the lattice $C_n(x,t)$
is expanded by the state with $k_\alpha$
\begin{eqnarray}
C_n(x,t) &=& \langle 0| \pi(x,t)\pi(0,t)\Omega_n|0\rangle\\
&=& \sum_{\alpha=0}^{N_\alpha} A_{n\alpha}(t) \phi_\alpha(x),
\end{eqnarray}
where $\phi_\alpha(x)$ corresponds to $\phi(x;k_\alpha)$ 
with discrete momenta
and $A_{n\alpha}(t) = B_{n\alpha} e^{-E_\alpha t}$ with 
$B_{n\alpha} = \langle \pi\pi;k_\alpha|\Omega_n|0\rangle$ and 
$E_\alpha^2 = 4 (m^2+k_\alpha^2)$.
We only consider elastic two-pion states.
$\Omega_n$ is a two-pion operator at the source ($t = 0$).
The different index $n$ denotes different operator $(n = 0, \cdots, N_\Omega)$, 
such as operators using different smearing.
$N_\alpha$ expresses the effective number of 
the states contributing to $C_n(x,t)$.
The number decreases as $t$ increases, because contributions of 
higher energy states are exponentially suppressed by $t$.

Using a function $f(t,m)$, which satisfies
\begin{equation}
f(t,m) A_{n\alpha}(t) = k^2_\alpha A_{n\alpha}(t),
\end{equation}
the sum of the reduced BS wave function is calculated from $C_n(x,t)$ as,
\begin{equation}
(\Delta + f(t,m))C_n(x,t) =
\sum_{\alpha=0}^{N_\alpha} A_{n\alpha}(t) h_\alpha(x) ,
\label{eq:thal_base}
\end{equation}
where $h_\alpha(x) = h(x;k_\alpha)$.

\subsection{Infinite term expansion}

The coefficient $V_i(x)$ is defined to be $k$ independent in 
the infinite term expansion of $h_\alpha(x)$~\cite{Aoki:2017yru} as,
\begin{equation}
h_\alpha(x) = \sum_{i=0}^\infty V_i(x) \Delta^i \phi_\alpha(x) .
\end{equation}
In this case, the right-hand side of Eq.~(\ref{eq:thal_base}) can be
expressed by $\Delta^i C_n(x,t)$ as,
\begin{eqnarray}
\sum_{\alpha=0}^{N_\alpha} A_{n\alpha}(t) h_\alpha(x)&=&
\sum_{\alpha=0}^{N_\alpha} A_{n\alpha}(t) \sum_{i=0}^\infty V_i(x) \Delta^i \phi_\alpha(x)\\
&=&
\sum_{i=0}^\infty V_i(x) \sum_{\alpha=0}^{N_\alpha} A_{n\alpha}(t) \Delta^i \phi_\alpha(x)
\label{eq:thal_1}\\
&=&
\sum_{i=0}^\infty V_i(x) \Delta^i C_n(x,t) .
\label{eq:thal_2}
\end{eqnarray}
In Eq.~(\ref{eq:thal_1}) summations for $\alpha$ and $n$ are exchanged 
thanks to $k$ independence of $V_i(x)$.
Since this is the ideal case using the infinite term expansion,
this discussion cannot be applicable to a practical calculation.

\subsection{Truncated expansion}

In the truncated expansion of $h_\alpha(x)$ with $N$ derivative terms,
the coefficients depend on $k_\alpha$~\cite{Yamazaki:2017gjl},
\begin{equation}
h_\alpha(x) = \sum_{i=0}^N V_{i\alpha}(x) \Delta^i \phi_\alpha(x) .
\end{equation}
Because of the $k_\alpha$ dependence of $V_{i\alpha}(x)$,
the summations for $n$ and $\alpha$ are not allowed to be
exchanged,
\begin{equation}
\sum_{\alpha=0}^{N_\alpha} A_{n\alpha}(t) \sum_{i=0}^N V_{i\alpha}(x) \Delta^i \phi_\alpha(x)
\ne
\sum_{i=0}^N V_i(x) \Delta^i C_n(x,t) ,
\label{eq:finv_eq}
\end{equation}
in contrast to the case of the infinite term expansion.
However, the time-dependent HALQCD method 
expresses the left hand side of 
Eq.~(\ref{eq:finv_eq})
by a similar form
to the infinite term expansion Eq.~(\ref{eq:thal_2})
with a coefficient $\overline{V}_i(x)$,
\begin{equation}
(\Delta + f(t,m))C_n(x,t) =
\sum_{i=0}^N \overline{V}_i(x) \Delta^i C_n(x,t) .
\end{equation}
It is apparent that
$\overline{V}_i(x) \ne V_i(x)$ from Eq.(\ref{eq:finv_eq}).
For convenience we define a matrix $M(x,t)$,
whose component $M_{ni}(x,t) = \Delta^i C_n(x,t)$.

When $N = N_\Omega$ 
(the numbers for $\overline{V}_i(x)$ and $C_n(x,t)$ are the same),
if the matrix $M(x,t)$ is a regular matrix, 
$M(x,t)$ has its inverse, and then $\overline{V}_i(x)$ is given as
\begin{equation}
\overline{V}_i(x) = \sum_{n=0}^N M_{in}^{-1}(x,t) (\Delta + f(t,m))C_n(x,t) .
\end{equation}
In order to understand $\overline{V}_i(x)$ determined from the equation,
we represent the equation by vectors and matrices as,
\begin{equation}
\overline{V}(x) = M^{-1}(x,t) A(t) h(x) ,
\end{equation}
where we use Eq.~(\ref{eq:thal_base}), and
the components for $\overline{V}(x), A(t), h(x)$
are $\overline{V}_i(x), A_{n\alpha}(t), h_\alpha(x)$, respectively.

In the case of $N_\alpha \ne N$
(the number of the states in $C_n(x,t)$ differs from 
the number of $\overline{V}_i(x)$),
the matrix $A(t)$ does not have
the inverse matrix, so that $M^{-1}(x,t)$ cannot be decomposed into
two inverse matrices $A^{-1}(t)$ and $\Phi^{-1}(x)$, where 
$\Phi_{\alpha i}(x)=\Delta^i\phi_\alpha(x)$,
albeit $M(x,t) = A(t) \Phi(x)$.
Therefore, $\overline{V}_i(x)$ is a function of $A_{n\alpha}(t), h_\alpha(x)$, 
and $\Delta^i \phi_\alpha(x)$.
It means that $\overline{V}_i(x)$ depends on the choice of the operators
to calculate $C_n(x,t)$.

If $A(t)$ is regular with $N_\alpha = N$, $A(t)$ has the inverse,
and then $M^{-1}(x,t) = \Phi^{-1}(x) A^{-1}(t)$.
In this case, the operator dependence of $\overline{V}_i(x)$ disappears,
\begin{equation}
\overline{V}(x) = M^{-1}(x,t) A(t) h(x)
= \Phi^{-1}(x) h(x).
\end{equation}
Although $\overline{V}(x)$ depends on $N_\alpha(=N)$ and also $k_\alpha$,
it gives the correct scattering amplitudes 
at only $k = k_\alpha$ by solving 
the Schr\"odinger equation~\cite{Aoki:2017yru}. 
The values of $k_\alpha$, however,
cannot be determined by the time-dependent HALQCD method.

In order to satisfy $N_\alpha = N$,
one needs to calculate $C_n(x,t)$ in a large $t$ region,
where contributions from higher energy states 
must be sufficiently suppressed compared to those from the states
of $\alpha = 0,\cdots,N_\alpha$.
It might be also possible to 
adopt appropriate operators which strongly couple to
the specific states.
It is the same condition to calculate the energy from 
the matrix of the time correlation function using 
the generalized eigenvalue problem~\cite{Luscher:1990ck},
or may be more severe,
because it must be satisfied in all $x$ for precise determination of the potential $V_{i\alpha}(x)$.

\section*{Acknowledgements}
This work is supported in part by Japan Society for the Promotion of Science (JSPS) Grants-in-Aid for Scientific Research for Young Scientists (A) No. 16H06002.

\bibliographystyle{apsrev4-1}
\bibliography{reference}

\end{document}